\RequirePackage{fixltx2e}
\documentclass[nofootinbib,reprint,showpacs,noeprint,
  prc,aps,10pt,superscriptaddress,altaffilletter,floatfix,longbibliography]{revtex4-1}
\pdfoutput=1
\usepackage{graphicx}
\usepackage{color}
\usepackage{hyperref}
\usepackage{epstopdf}
\usepackage{amsmath}
\usepackage{amssymb}
\usepackage[capitalise]{cleveref}
\usepackage{ifthen}
\usepackage{multirow}
\usepackage{booktabs}
\usepackage{dcolumn}
\usepackage{array}
\usepackage{textcomp}
\newcolumntype{C}[1]{>{\centering\let\newline\\\arraybackslash\hspace{0pt}}m{#1}}

\graphicspath{{figs/}{/}}

\begin{document}

\newcommand{\isot}[2]{$^{#2}$#1}
\newcommand{\isotbold}[2]{$^{\boldsymbol{#2}}$#1}
\newcommand{\xeiso}{\isot{Xe}{136}}
\newcommand{\thsrc}{\isot{Th}{228}}
\newcommand{\cosrc}{\isot{Co}{60}}
\newcommand{\rasrc}{\isot{Ra}{226}}
\newcommand{\cssrc}{\isot{Cs}{137}}
\newcommand{\xeprob}{\isot{Xe}{137}}
\newcommand{\thch}{\isot{Th}{232}}
\newcommand{\uch}{\isot{U}{238}}
\newcommand{\betascale}  {$\beta$-scale}
\newcommand{\kevkgyr}  {keV$^{-1}$ kg$^{-1}$ yr$^{-1}$}
\newcommand{\nonubb}  {$0\nu \beta \beta$}
\newcommand{\nonubbbf}  {$\boldsymbol{0\nu \beta \beta}$}
\newcommand{\twonubb} {$2\nu \beta \beta$}
\newcommand{\vadc} {ADC$_\text{V}$}
\newcommand{\uadc} {ADC$_\text{U}$}
\newcommand{\mus} {\textmu{}s}
\newcommand{\chisq} {$\chi^2$}
\newcommand{\mum} {\textmu{}m}
\newcommand{\checkit}[1]{{\color{red}#1}}
\newcommand{\RunTwoA}{Run 2a}
\newcommand{\RunTwo}{Run 2}
\newcommand{\RunTwoBC}{Runs 2b and 2c}
\newcommand{\SP}[1]{\textsuperscript{#1}}
\newcommand{\SB}[1]{\textsubscript{#1}}
\newcommand{\SPSB}[2]{\rlap{\textsuperscript{#1}}\SB{#2}}
\newcommand{\pmasy}[3]{#1\SPSB{$+$#2}{$-$#3}}
\newcommand{\matel}{$M^{2\nu}$}
\newcommand{\psfac}{$G^{2\nu}$}
\newcommand{\tbeta}{T$_{1/2}^{0\nu\beta\beta}$}
\newcommand{\exolimit}[1][true]{\pmasy{2.6}{1.8}{2.1}$ \cdot 10^{25}$}
\newcommand{\exomeasurement}{\tbeta{}= \exolimit{}~yr}
\newcommand{\U}{\text{U}}
\newcommand{\V}{\text{V}}
\newcommand{\X}{\text{X}}
\newcommand{\Y}{\text{Y}}
\newcommand{\Z}{\text{Z}}
\newcommand{\bqcm}{${\rm Bq~m}^{-3}$}
\newcommand{\nonunorm}{N_{{\rm Err, } 0\nu\beta\beta}}
\newcommand{\nonunum}{n_{0\nu\beta\beta}}
\newcommand{\cussim}[1]{$\sim$#1}
\newcommand{\halflife}[1]{$#1\cdot10^{25}$~yr}
\newcommand{\numspec}[3]{$N_{^{#2}\mathrm{#1}}=#3$}
\newcommand{\ran}[2]{
  \newcommand\myvar{#1}
  \if\myvar0
    $<#2$
  \else
    #1~\textendash{}~#2
  \fi
}
\newcommand{\Tmu}{TPC muon}
\newcommand{\Tmus}{TPC muons}
\newcommand{\bipo}{Bi\textendash{}Po events}
\newcommand{\teflon}{Teflon\textsuperscript{\textregistered}}

\title{Investigation of radioactivity-induced backgrounds in EXO-200}

\newcommand{\Alabama}{\affiliation{Department of Physics and Astronomy, University of Alabama, Tuscaloosa, Alabama 35487, USA}}
\newcommand{\Alberta}{\affiliation{University of Alberta, Edmonton, Alberta, Canada}}
\newcommand{\Bern}{\affiliation{LHEP, Albert Einstein Center, University of Bern, Bern, Switzerland}}
\newcommand{\CALTECH}{\affiliation{Kellogg Lab, Caltech, Pasadena, California 91125, USA}}
\newcommand{\Carleton}{\affiliation{Physics Department, Carleton University, Ottawa, Ontario K1S 5B6, Canada}}
\newcommand{\CSU}{\affiliation{Physics Department, Colorado State University, Fort Collins, Colorado 80523, USA}}
\newcommand{\Drexel}{\affiliation{Department of Physics, Drexel University, Philadelphia, Pennsylvania 19104, USA}}
\newcommand{\Duke}{\affiliation{Department of Physics, Duke University, and Triangle Universities Nuclear Laboratory (TUNL), Durham, North Carolina 27708, USA}}
\newcommand{\IBS}{\affiliation{IBS Center for Underground Physics, Daejeon, Korea}}
\newcommand{\IHEP}{\affiliation{Institute of High Energy Physics, Beijing, China}}
\newcommand{\Illinois}{\affiliation{Physics Department, University of Illinois, Urbana-Champaign, Illinois 61801, USA}}
\newcommand{\Indiana}{\affiliation{Physics Department and CEEM, Indiana University, Bloomington, Indiana 47405, USA}}
\newcommand{\ITEP}{\affiliation{Institute for Theoretical and Experimental Physics, Moscow, Russia}}
\newcommand{\Laurentian}{\affiliation{Department of Physics, Laurentian University, Sudbury, Ontario P3E 2C6, Canada}}
\newcommand{\Maryland}{\affiliation{Physics Department, University of Maryland, College Park, Maryland 20742, USA}}
\newcommand{\Munich}{\affiliation{Technische Universit\"at M\"unchen, Physikdepartment and Excellence Cluster Universe, Garching 80805, Germany}}
\newcommand{\SDakota}{\affiliation{Physics Department, University of South Dakota, Vermillion, South Dakota 57069, USA}}
\newcommand{\Seoul}{\affiliation{Department of Physics, University of Seoul, Seoul, Korea}}
\newcommand{\SLAC}{\affiliation{SLAC National Accelerator Laboratory, Stanford, California 94025, USA}}
\newcommand{\Stanford}{\affiliation{Physics Department, Stanford University, Stanford, California 94305, USA}}
\newcommand{\Stony}{\affiliation{Department of Physics and Astronomy, Stony Brook University, SUNY, Stony Brook, New York 11794, USA}}
\newcommand{\TRIUMF}{\affiliation{TRIUMF, Vancouver, BC, Canada}}
\newcommand{\UMass}{\affiliation{Amherst Center for Fundamental Interactions and Physics Department, University of Massachusetts, Amherst, MA 01003, USA}}
\newcommand{\WIPP}{\affiliation{Waste Isolation Pilot Plant, Carlsbad, New Mexico 88220, USA}}
\author{J.B.~Albert}\Indiana
\author{D.J.~Auty}\altaffiliation{Present address: University of Alberta, Edmonton, AB, Canada}\Alabama
\author{P.S.~Barbeau}\Duke
\author{D.~Beck}\Illinois
\author{V.~Belov}\ITEP
\author{C.~Benitez-Medina}\altaffiliation{Present address: Intel Corporation, Hillsboro, OR, USA}\CSU
\author{M.~Breidenbach}\SLAC
\author{T.~Brunner}\Stanford
\author{A.~Burenkov}\ITEP
\author{G.F.~Cao}\IHEP
\author{C.~Chambers}\CSU
\author{B.~Cleveland}\altaffiliation{Also SNOLAB, Sudbury, ON, Canada}\Laurentian
\author{M.~Coon}\Illinois
\author{A.~Craycraft}\CSU
\author{T.~Daniels}\SLAC
\author{M.~Danilov}\ITEP
\author{S.J.~Daugherty}\Indiana
\author{C.G.~Davis}\altaffiliation{Present address: Naval Research Lab, Washington D.C., USA}\Maryland
\author{J.~Davis}\SLAC
\author{S.~Delaquis}\Bern
\author{A.~Der Mesrobian-Kabakian}\Laurentian
\author{R.~DeVoe}\Stanford
\author{T.~Didberidze}\Alabama
\author{A.~Dolgolenko}\ITEP
\author{M.J.~Dolinski}\Drexel
\author{M.~Dunford}\Carleton
\author{W.~Fairbank Jr.}\CSU
\author{J.~Farine}\Laurentian
\author{W.~Feldmeier}\Munich
\author{P.~Fierlinger}\Munich
\author{D.~Fudenberg}\Stanford
\author{G.~Giroux}\altaffiliation{Present address: Dept.\ of Physics, Queen's University, Kingston, ON, Canada}\Bern
\author{R.~Gornea}\Bern
\author{K.~Graham}\Carleton
\author{G.~Gratta}\Stanford
\author{C.~Hall}\Maryland
\author{S.~Herrin}\SLAC
\author{M.~Hughes}\Alabama
\author{M.J.~Jewell}\Stanford
\author{X.S.~Jiang}\IHEP
\author{A.~Johnson}\SLAC
\author{T.N.~Johnson}\Indiana
\author{S.~Johnston}\UMass
\author{A.~Karelin}\ITEP
\author{L.J.~Kaufman}\Indiana
\author{R.~Killick}\Carleton
\author{T.~Koffas}\Carleton
\author{S.~Kravitz}\Stanford
\author{A.~Kuchenkov}\ITEP
\author{K.S.~Kumar}\Stony
\author{D.S.~Leonard}\IBS
\author{C.~Licciardi}\Carleton
\author{Y.H.~Lin}\Drexel
\author{J.~Ling}\Illinois
\author{R.~MacLellan}\SDakota
\author{M.G.~Marino}\email[Corresponding author: ]{michael.marino@mytum.de}\Munich
\author{B.~Mong}\Laurentian
\author{D.~Moore}\Stanford
\author{R.~Nelson}\WIPP
\author{A.~Odian}\SLAC
\author{I.~Ostrovskiy}\Stanford
\author{A.~Piepke}\Alabama
\author{A.~Pocar}\UMass
\author{C.Y.~Prescott}\SLAC
\author{A.~Rivas}\SLAC
\author{P.C.~Rowson}\SLAC
\author{J.J.~Russell}\SLAC
\author{A.~Schubert}\Stanford
\author{D.~Sinclair}\TRIUMF\Carleton
\author{E.~Smith}\Drexel
\author{V.~Stekhanov}\ITEP
\author{M.~Tarka}\Stony
\author{T.~Tolba}\Bern
\author{R.~Tsang}\Alabama
\author{K.~Twelker}\Stanford
\author{J.-L.~Vuilleumier}\Bern
\author{A.~Waite}\SLAC
\author{J.~Walton}\Illinois
\author{T.~Walton}\CSU
\author{M.~Weber}\Stanford
\author{L.J.~Wen}\IHEP
\author{U.~Wichoski}\Laurentian
\author{J.~Wood}\WIPP
\author{L.~Yang}\Illinois
\author{Y.-R.~Yen}\Drexel
\author{O.Ya.~Zeldovich}\ITEP

\collaboration{EXO-200 Collaboration}
\noaffiliation

\date{\today}

\begin{abstract}

The search for neutrinoless double-beta decay (\nonubb{}) requires extremely
low background and a good understanding of their sources and
their influence on the rate in the region of parameter space relevant to the \nonubb{} signal.
We report on studies of various $\beta$- and $\gamma$-backgrounds in the
liquid-xenon-based EXO-200 \nonubb{} experiment.  With this work we try to better
understand the location and strength of specific background sources and
compare the conclusions to radioassay results taken before and during detector construction.  Finally, we
discuss the implications of these studies for EXO-200 as well as for the
next-generation, tonne-scale nEXO detector. 

\end{abstract}


\maketitle

\section{Introduction}\label{sec:Intro}

Neutrinoless double-beta decay (\nonubb{}) is a hypothetical process that may
be observable in even-even nuclei in which normal $\beta$-decay is either
energetically disallowed or highly forbidden.
The discovery
of this lepton-number-violating decay would indicate that the neutrino is a
massive Majorana particle~\cite{Schechter:1981bd}.  Current best limits from
\isot{Xe}{136}~\cite{Gando:2012zm,Nature2014} and 
\isot{Ge}{76}~\cite{Agostini:2013mzu}
require the \nonubb{} half-life
to be greater than \cussim{$10^{25}$~y}, posing significant experimental
challenges for experiments seeking to improve detector reach.  
The challenge of identifying and rejecting backgrounds at the low
energies of interest necessitates the use of radio-clean detector materials,
passive and active shielding, and sophisticated analysis techniques (e.g.\
signal pattern recognition).  

The background expectation, including the predicted counts 
detected in the \nonubb{} signal region and the event distribution arising from internal and external
background sources, is essential both for the optimization of detector design
prior to construction, as well as for the analysis and interpretation of data.
Only a quantitative understanding of the background expectation value, based on
various observables, allows the identification of a possible event excess with the
existence of a \nonubb{} signal.  
We present investigations on $\beta$- and $\gamma$-backgrounds to the EXO-200
\nonubb{} experiment, comparing current results with those produced prior to
detector operation and discussing the implications both for the continued
running of the EXO-200 detector as well as for the proposed, tonne-scale nEXO
detector. 

\begin{figure}
\includegraphics[width=0.98\columnwidth]{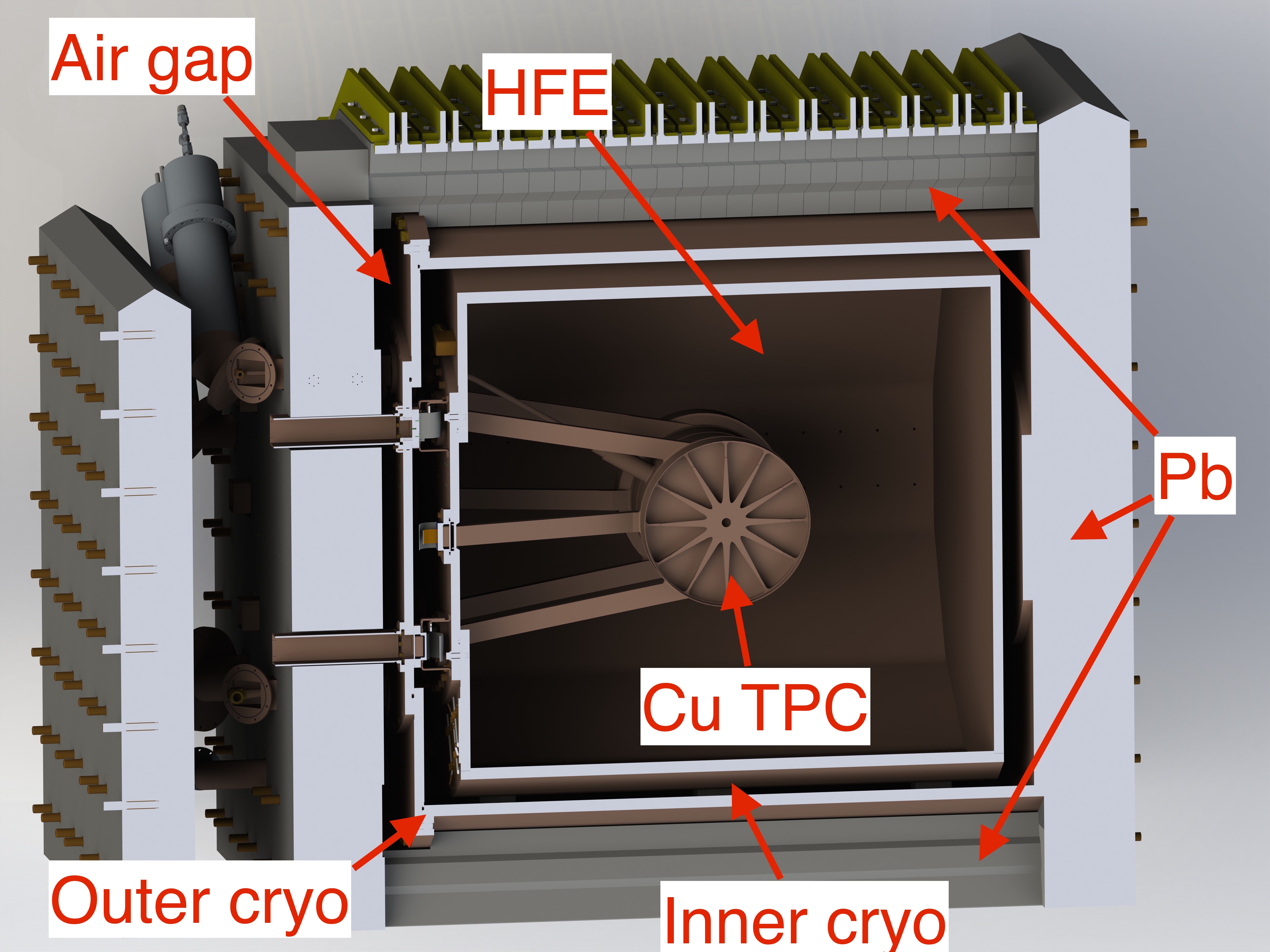}
\caption{(Color online) Cutaway engineering drawing of the EXO-200 detector
showing the relationship between  the TPC, HFE, Cu cryostats and Pb shield.  The TPC,
mounted on the wall of the inner cryostat with six copper supports, is surrounded by HFE
cryogenic fluid.  A vacuum layer lies between the inner and outer cryostats, and an
air layer (`Air gap') between the outer cryostat and the Pb shield.  The clean
room and surrounding active muon veto detectors are omitted for clarity. 
}
\label{fig:DetectorContext}
\end{figure}

\begin{figure}
\includegraphics[width=0.98\columnwidth]{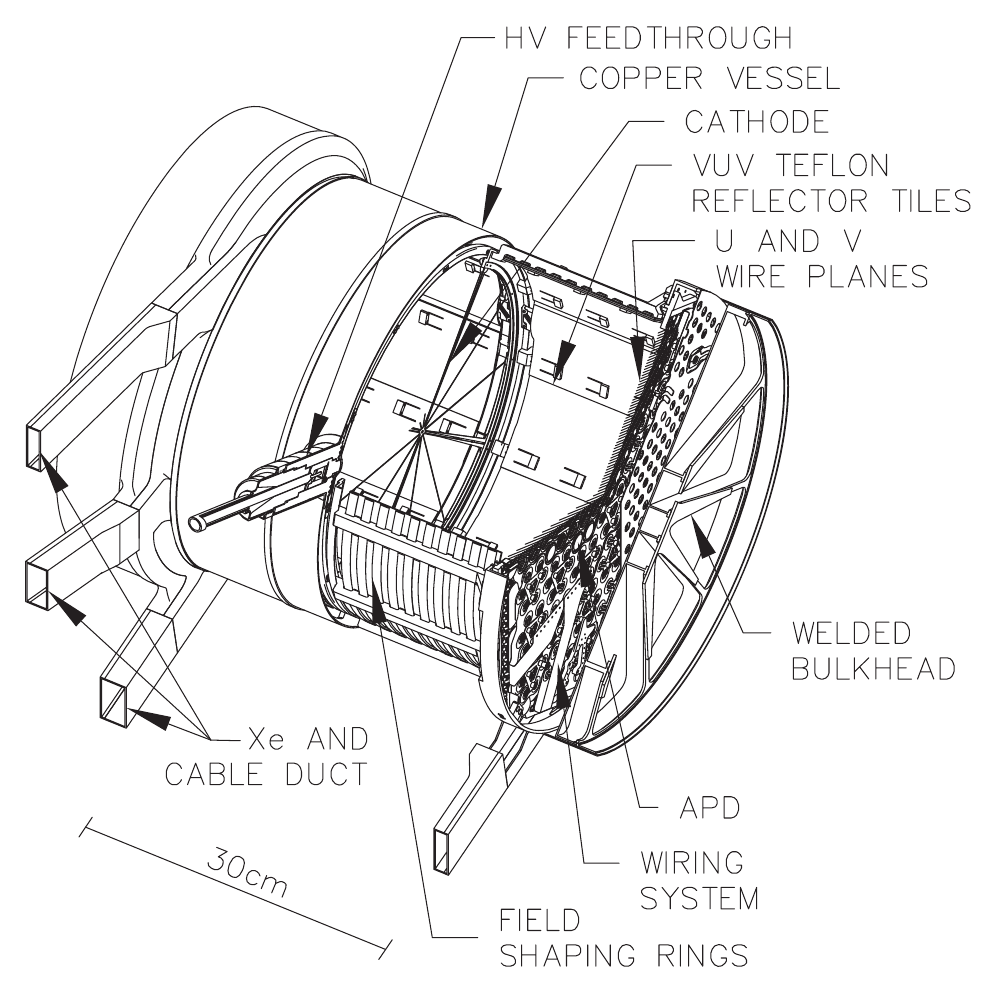}
\caption{Cutaway view of the EXO-200 Cu TPC.  `Wiring system' refers to the
flat polyimide cables behind the APD platter. Additional flat polyimide cables are
within the Cu TPC legs (`Cable duct' in figure).  Acrylic spacers, insulators and field
cage resistors are components of the as shown `Field shaping rings' system, and
are all external to the \teflon{} reflectors (`VUV (Vacuum Ultraviolet) Teflon Reflector Tiles').  A Cu tube and
support system used for deployment of calibration sources outside the Cu
vessel is not shown. }
\label{fig:DetectorSchematic}
\end{figure}

A comprehensive description of EXO-200 may be found
in~\cite{Auger:2012gs}.  The detector is a cylindrical liquid xenon
(LXe) time projection chamber (TPC), roughly 40~cm in diameter and 44~cm in
length.  The LXe is enriched to 80.6\% in \isot{Xe}{136}, the \nonubb{}
candidate  ($Q=2457.83\pm 0.37$~keV~\cite{Redshaw:2007}).  Energy depositions
in the xenon produce both scintillation light and charge.  The scintillation
light is detected by 468~avalanche photodiodes (APDs) on both ends of the TPC.  The
electric field produced by the negatively biased cathode at the center of the
TPC sweeps the ionization electrons to the two ends, where they are collected on wire
planes.  Cu rings, supported by acrylic spacers, form the field cage
surrounding the drift region, assuring a uniform electric field.  Inside of this cage is a cylindrical \teflon{}
light reflector.  The two wire planes at either end of the TPC are each composed of
collection (U) wires and induction (V) wires crossed at 60~degrees.  The
signals from both U- and V-wires are read out.  The APDs are situated behind the U- and V-wires, resulting in good light transmission.  Behind the APD mounts on each
side of the TPC are flat polyimide signal-readout cables, which are separated from the Cu
bulkhead by a layer of \teflon{}.  A Cu tube assembly used for the deployment
of calibration sources is located just outside the TPC.  A more detailed view
of the components in and around the TPC may be seen
in~\cref{fig:DetectorContext,fig:DetectorSchematic}. 

The TPC is surrounded on all sides and kept at \cussim{167~K} by \cussim{50~cm} of HFE-7000
cryofluid~\cite{3m} (HFE) that is housed inside a double-walled vacuum-insulated Cu
cryostat.  Six copper supports (`TPC legs') mount the TPC to the removable hatch of the cryostat and 
provide a conduit for additional signal cables.  The HFE further acts as a shield
from external radioactivity.  A Pb shield of \cussim{25~cm} thickness, located outside the Cu cryostat,
provides additional shielding.  This apparatus is in
a cleanroom located underground at a depth of $1585^{+11}_{-\phantom{0}6}$~meters water
equivalent~\cite{Esch:2004zj} at the Waste Isolation Pilot Plant (WIPP) near
Carlsbad, NM, USA. Four of the six sides of the cleanroom are instrumented with
large plastic scintillator panels, providing $96.0\pm0.5\%$ coverage of
muon-associated events detected in the TPC.  This efficiency is determined
by measuring the percentage of muons passing through the TPC (events tagged at
$>99.9$\% efficiency) that also correspond to muon veto panel events.

The data analysis methods used in EXO-200 are described  
in~\cite{Albert2013}.  Detector events are reconstructed to determine: (1)
event multiplicity, (2) event position, and (3) total event energy.  The
`event multiplicity' refers to the number of charge deposits reconstructed for a
particular event; `single-site' (SS) is used to refer to events with only one
discernible charge deposit (with a characteristic radius of \cussim{2-3~mm}) and
`multi-site' (MS) to those with two or more.  This quantity allows powerful
discrimination between signal and background: in the EXO-200 detector,
\nonubb{} events are \cussim{90\%~SS}, whereas $\gamma$s with energies around
the $\beta\beta$ Q-value are 20-30\% SS.  The event position is determined by
combining information from the APDs and U-/V-wires.  The total energy of the
event is measured in both the charge and light channels, which are then
linearly combined to improve the energy resolution~\cite{Conti2003}.  

In the following sections, unless otherwise specified, we present results using
`low-background data', or data passing all data quality selection requirements,
corresponding to $477.60\pm 0.01$~days of live-time accumulated in the period
from September 22, 2011 to September 1, 2013 (\RunTwo{}).  Limits on the
\nonubb{} decay of \isot{Xe}{136} using these data have been
presented in Ref.~\cite{Nature2014}. 

To search for a \nonubb{} signal in the data, a fit is performed minimizing a
negative log-likelihood function constructed using a signal and background
model composed of probability density functions (PDFs) from simulation.  The
fit proceeds over both multiplicity categories (SS and MS) and uses the
observables energy and `standoff distance' (SD), the latter a parameterization of the
position of an event in the detector.  The background model is composed of the
combination of all expected background sources contributing a non-negligible
number of counts to the data.  The
list of sources is based on information from assays of the materials
used in the detector construction.  These, combined with results from simulation,
provide an estimate of the expected strengths of contributions from various sources and
locations.  In principle, this set includes a large number of different
components for each separate detector piece, but in practice the shapes of the
energy and SD distributions for many of these are degenerate and therefore
indistinguishable in the data.  For this reason, we include only the following
components in our background model:

\begin{itemize}
  \item Cu TPC vessel: \isot{K}{40}, \isot{Mn}{54}, \cosrc{}, \isot{Zn}{65}, \thch{}, and \uch{}
  \item LXe: \isot{Xe}{135}, \isot{Xe}{137} and \isot{Rn}{222}, 
  \item Air gap between Cu cryostat and Pb shield: \isot{Rn}{222}
  \item `Far-source' \isot{Th}{232} (in Cu cryostat),
  \item neutron-capture related: \isot{Xe}{136} neutron capture in the LXe,
\isot{H}{1} neutron-capture in the HFE, and \isot{Cu}{63},\isot{Cu}{65} neutron
capture in Cu components (LXe vessel, inner and outer cryostats)
\end{itemize}

Note that, because of the aforementioned degeneracies, the Cu TPC vessel PDF
includes contributions from background sources at a similar radius (i.e.\ APDs,
signal cables, Cu calibration tube etc.).  The inclusion of \isot{Rn}{222} in
the air gap between the Cu cryostat and Pb shield accounts for a far-source of
\isot{U}{238}.  \isot{Rn}{220} is not included because of its short half-life
and because measurements of the clean-room air indicated only negligible amounts  
in comparison to \isot{Rn}{222}.
For construction of the PDFs, the
\thch{} and \uch{} chains are assumed to be in secular equilibrium; we find no
evidence from the data that this assumption is unwarranted.   The component fit
degeneracy is not important for the computation of the background expectation
value as different assignments would lead to the same background rate. However,
our ability to unfold quantitative values for the radioactivity content of
these components is limited by this.

In the following sections, we analyze the predictions from assays for \thch{}
and \uch{} and compare to measurements from data, focusing on the the $\pm2\sigma$ energy range
around the \nonubb{} Q-value (ROI).  Using low-background data, we investigate the placement of
\thch{} and \uch{} and look at the time behavior of different backgrounds over
the \RunTwo{} time period.  We look for evidence of \isot{Xe}{137}
production, and present results on the presence of \isot{Rn}{222} in the LXe
and in air outside the cryostat.   

\begin{table} 
\renewcommand{\arraystretch}{1.3}

\caption{Expected SS event rates (90\% CL) in the $\pm2\sigma$ \nonubb{} ROI
(with $\sigma/E=1.53\%$~\cite{Nature2014}) from \thch{} and
\uch{} backgrounds from different detector components.  These have been calculated using the
Geant4-based~\cite{GEANT42006} Monte Carlo simulation~\cite{Albert2013} of the
EXO-200 detector and from assay results~\cite{Leonard2008490}.  An exception is `Pb shielding',
whose results presented in this table are from Ref.~\cite{Auger:2012gs}.  See text and
\cref{fig:DetectorContext,fig:DetectorSchematic} for more details on the
location of different components.  For assay results producing a measurement of
activity (as opposed to an upper limit), the 90\%~CL range of expected event rates are
given. 
The lower (upper) limit of the total predicted rate is produced by summing the lower (upper) limits of each component.  This results in a conservative bracketing of the expectation values.
A summary
of previous results presented in~\cite{Auger:2012gs} and produced with an
earlier Geant3-based~\cite{Geant3} version of the EXO-200 simulation software
are shown for comparison.  These previous results used a slightly larger ROI
($\sigma/E=1.6\%$). The rates for backgrounds from \thch{} and \uch{} estimated
from the fit to low-background data~\cite{Nature2014} are also given
for comparison.  
}

\label{tab:ExpBkgdCts}
\begin{ruledtabular}
\begin{tabular}{>{\hangindent=0.02\columnwidth}p{0.49\columnwidth}cc}
    \toprule
	Part/Material & \multicolumn{2}{m{0.47\columnwidth}}{\centering Expected counts in \nonubb{} ROI (Cts / yr)} \\
    & \thch{} & \uch{} \\
    \colrule

APDs                        &   \ran{0}{0.6}            &  \ran{0}{0.05}         \\  
Bronze cathode              &   $(6-10)\cdot10^{-5}$    &  $(6-8)\cdot10^{-3}$ \\
 Bronze wires                &   \ran{0.04}{0.07}        &  \ran{0.07}{0.08}  \\  
 Other Bronze                &   \ran{0.08}{0.13}        &  \ran{0.25}{0.3}   \\  
 Flat cables (in TPC)            &   \ran{0}{0.44}           &  \ran{1.5}{2.3}    \\  
 Flat cables (in TPC Legs)       &   \ran{0}{0.04}           &  \ran{0.1}{0.14}   \\  
 \teflon{} reflectors, HV insulators        &   \ran{0.06}{0.1}         &  \ran{0}{0.07}         \\  
 \teflon{} behind APDs           &   \ran{0.2}{0.3}          &  \ran{0}{0.08}         \\  
 Acrylic spacers, insulators &   \ran{0}{0.3}            &  \ran{0}{0.6}      \\  
 Field cage resistors        &   \ran{0}{0.01}           &  \ran{0}{0.01}     \\  
 Cu TPC vessel               &   \ran{0}{3}              &  \ran{0}{3.8}      \\  
 Cu TPC legs                 &   \ran{0}{0.07}           &  \ran{0}{0.05}         \\  
 HV cable                    &   \ran{0}{0.13}           &  \ran{0}{0.6}      \\
 Cu calibration tube + support &             \ran{0.1}{0.2}      &  \ran{0.15}{0.35}  \\
 Cu cryostat                 &   \ran{0}{1.3}            &  \ran{0}{0.5}          \\
 HFE-7000                    &   \ran{0}{0.1}            &  \ran{0}{0.1}      \\
 Pb shielding                &   \ran{0}{0.9}            &  \ran{0}{0.5}      \\
\colrule
\multicolumn{1}{m{0.47\columnwidth}}{Total predicted rates (sum):}                &   \ran{0.5}{7.7}          &  \ran{2}{9.5}
\\
\\
\multicolumn{1}{m{0.47\columnwidth}}{Total predicted rates from Ref.~\cite{Auger:2012gs}:}        &   \ran{0.9}{10.3}         & \ran{6.3}{26.8}
\\
\multicolumn{1}{m{0.47\columnwidth}}{Observed rates from Ref.~\cite{Nature2014}:} & \ran{10.3}{13.9} & \ran{5.3}{7.1} \\

\end{tabular}
\end{ruledtabular}
\end{table}

\begin{table*} 
\renewcommand{\arraystretch}{1.3}

\caption{Upper limits (90\% CL) on detector material activity (mBq) estimated using
low-background data (`This meas.' columns)--- see text for more details.  See
\cref{fig:DetectorContext,fig:DetectorSchematic} for a visualization of the
placement of different components.  Several sets of components share similar
positions in the detector construction.  These produce degenerate signals in
the detector and therefore have the same absolute activity limits.  Assay
results from Ref.~\cite{Auger:2012gs} and neutron activation measurements of the \teflon{} \thch{} and \uch{} surface activity are included for
comparison (`Assay' columns).  Limits from Ref.~\cite{Auger:2012gs} are at 90\%CL.  The error
bars for assays producing a measurement of activity (instead of a limit) in
Ref.~\cite{Auger:2012gs} are $\pm1\sigma$ and have been expanded here to 90\%
CL. Calculations for \isot{K}{40} for a number of small components were not
performed.  }

\label{tab:ExpLimits}
\newcommand{\indentme}{\setlength\parindent{2em}\setlength\hangindent{2em}}
\begin{ruledtabular}
\begin{tabular}{>{\hangindent=0.02\columnwidth}p{0.15\textwidth}ccp{0.05\textwidth}cp{0.05\textwidth}cp{0.05\textwidth}}
    \toprule
								& 					& \multicolumn{6}{c}{Detector material activity (mBq) }\\ 
\multirow{3}{*}{Part/Material} 	& \multirow{3}{*}{Quantity} & \multicolumn{2}{c}{\isot{K}{40}} 	&  \multicolumn{2}{c}{\thch{}} 		&  \multicolumn{2}{c}{\uch{}} \\ 
   							&				&	Assay 			& This meas.  					&	Assay 			& This meas. 		&	Assay 			& This meas. \\
    \colrule
APDs 						&	518 units 		&	$<0.13$ 			&	$<31 			$	&	$<0.09$         	&	$<1.0$ 	&	$<0.011$ 			&	$<1.7  $\\
Bronze cathode 				&	0.010~kg 		&	$<0.019$			&	---  	 				&	$0.00108\pm0.00019$	&	$<12$ 	&	$0.00364\pm0.00021$ &	$<9.7  $\\
Bronze wires 					&	0.083~kg 		&	$<0.16$ 			&	$<51 			$	&	$0.0090\pm0.0015$ 	&	$<1.9$ 	&	$0.0302\pm0.0017$ 	&	$<2.4  $\\
Other Bronze 					&	0.314~kg 		&	$<0.6$			&	$<51 			$	&	$0.0176\pm0.0027$ 	&	$<1.9$ 	&	$0.107\pm0.006$ 	&	$<2.4  $\\
Flat cables (TPC) 				&	7406~cm$^2$ 	&	$<0.9$			&	$<31 			$	&	$<0.07$       		&	$<1.5$ 	&	$0.43\pm0.06$		&	$<1.7  $\\
Flat cables (TPC Legs) 			&	10825~cm$^2$ &	$<1.4$			&	$<840			$	&	$<0.07$       		&	$<23$	&	$0.76\pm0.09$		&	$<36   $\\
\teflon{} (surface) 				\\
\indentme{}behind APDs, HV insulators 	&	43400~cm$^{2}$  &	--- 				&	--- 					&	$0.040\pm0.004$	&	$<1.5$ 	&	$<0.113$			&	$<1.66 $ \\
\indentme{}reflectors 		& 1660~cm$^{2}$ 	&	--- 				&	---					&	$0.006\pm0.001$     	&	$<0.73$	&	$0.004\pm0.001$			&	$<0.84 $ \\
Acrylic spacers, insulators 		&	1.460~kg		&	$<0.14$ 			&	---					&	$<0.024$      		&	$<0.73$	&	$<0.07$ 			&	$<0.84 $ \\
Field cage resistors				&	20~units 	 	&	$<0.08$ 			&	---					&	$<0.0006$     		&	$<0.73$	&	$<0.0017$ 			&	$<0.84 $ \\
Cu TPC vessel					&	32.736 kg 	 	&	$<60$ 			&	$<51 			$	&	$<0.5$        		&	$<1.9 $	&	$<1.5$ 				&	$<2.4  $\\
Cu TPC legs 					&	6.944~kg 		&	$<12$ 			&	$<1000		$		&	$<0.11$       		&	$<30$ 	&	$<0.33$				&	$<50   $\\
HV cable 						&	0.091~kg 		&	$<5$ 			&	$<95 			$	&	$<0.036$      		&	$<3.0$ 	&	$<0.6$ 				&	$<4.2  $\\
Cu calibration tube 				&	0.473~kg 		&	$<8$ 			&	$<126			$	&	$0.016\pm0.003$		&	$<4.8$ 	&	$0.043\pm0.001$		&	$<6.5  $\\
Cu wire calibration tube support	&	0.144~kg		&	$<11$ 			&	$<126			$	&	$0.027\pm0.002$		&	$<4.8$ 	&	$0.19\pm0.06$ 		&	$<5.5  $\\
Cu cryostat 					&	5901~kg 		&	$<72$ 			&	$<7.0\times10^5$		&	$<19$          		&	$<6500$	&	$<58$ 				&	$<18000$  \\
HFE-7000 					&	4140~kg 		&	$<20$ 			&	$<1200 	$			&	$<0.25$        		&	$<40$ 	&	$<0.8$				&	$<57   $

\end{tabular}
\end{ruledtabular}

\end{table*}

\section{Background expectations} 

We have used results from assays of \thch{} and \uch{} in detector materials
to compute the expected counts in the \nonubb{} ROI, defined as
the measured activity (mBq, from \cite{Leonard2008490,Auger:2012gs}) multiplied
by the efficiency calculated from the EXO-200 Geant4-based~\cite{GEANT42006}
Monte Carlo simulation package (MC), see e.g.~\cite{Albert2013}.  Most of the
assay results used in this work were at hand before construction
\cite{Leonard2008490}, with the exception of new neutron activation results of the
\teflon{} used in detector construction.  These new results indicated a higher total
expected \thch{} activity of $46\pm4~\mu$Bq (attributed to activity on the \teflon{} surface) than the limits presented in
Ref.~\cite{Auger:2012gs}.
Similar estimates of counts in the \nonubb{} ROI were presented previously
in~\cite{Auger:2012gs} with an earlier Geant3-based~\cite{Geant3} version of
the simulation software.  (The newer simulation also includes more detailed
geometries and more advanced electronic signal generation.) 
The summary of these are shown together with the
current results in \cref{tab:ExpBkgdCts}.  In addition, we compare the total
expected counts in the ROI for \thch{} and \uch{} to the results from a
maximum-likelihood fit to the low-background data~\cite{Nature2014}.  

Ref.~\cite{Auger:2012gs} also included results for \isot{K}{40} due to its
relevance as a background for \twonubb{}, but we omit these in the presentation
of \cref{tab:ExpBkgdCts} here because they do not contribute to the \nonubb{} ROI.  The results of
the low-background fit are
however consistent with estimates of the contributions of \isot{K}{40}. 

The results from the pre- and post-data-taking simulations are in good agreement, although the current
numbers are somewhat lower.  This can be understood as owing to a better
measured resolution at the $\beta\beta$ Q-value ($\sigma/E=1.53\%$~\cite{Nature2014})
and improvements in the simulation software including more realistic geometries and a better
estimation of the SS/MS discrimination.  The predicted rates from assay and the observed rates
of \thch{} and \uch{} from the fit to low-background data also show reasonable
agreement, although the prediction for \thch{} backgrounds is slightly lower than
that observed in the data.

Results from~\cite{Nature2014} indicate that \isot{Xe}{137} is another
important background in the \nonubb{} ROI.  This short-lived species ($\tau =
229.1\pm0.8$~s~\cite{Browne20072173}) is produced by neutron capture on
\isot{Xe}{136} in LXe and estimates from simulation of its production in the LXe
are consistent with the experimental results.  Here we present an analysis
searching for evidence of the production of \isot{Xe}{137} in coincidence with
cosmic-ray-muon events.  More details concerning the production of
\isot{Xe}{137} and the detection of the associated radiative neutron capture 
as well as the impact of other cosmogenic backgrounds to the search for
\nonubb{} with the EXO-200 detector will be published separately.

We have explored the possibility that better constraints on intrinsic
activities in detector components may be produced by using the low-background data.  In other
words: given the observed rates, we can derive an estimate on the activities in
detector components.  This uses conservative assumptions, specifically that
all the counts assigned to a particular PDF \emph{type} in the low-background
fit (e.g.\ \isot{K}{40}, \thch{}, \uch{}) are coming exclusively from one
component.  
This approach lifts the fit component degeneracy at the expense of sensitivity.
By construction it yields only limits as no fitting is performed.
For almost all components, this produces weaker limits
than the assay.  Exceptions are \isot{K}{40} and \uch{} in the Cu
TPC vessel, where the results from low-background data are more constraining than, or very
similar to, those from assay.  Results from low-background data
yield limits on \isot{K}{40} (\uch{}) in the Cu TPC vessel of $<1.6$~mBq/kg
($<0.07$~mBq/kg), which one can compare to results from assay: $<1.8$~mBq/kg
($<0.05$~mBq/kg).  Full results from this study are presented in
\cref{tab:ExpLimits}.   

\section{Investigations of background components}

\subsection{Location of \texorpdfstring{\thch{}/\uch{}}{Th-232/U-238}}

\begin{figure}
\includegraphics[width=0.98\columnwidth]{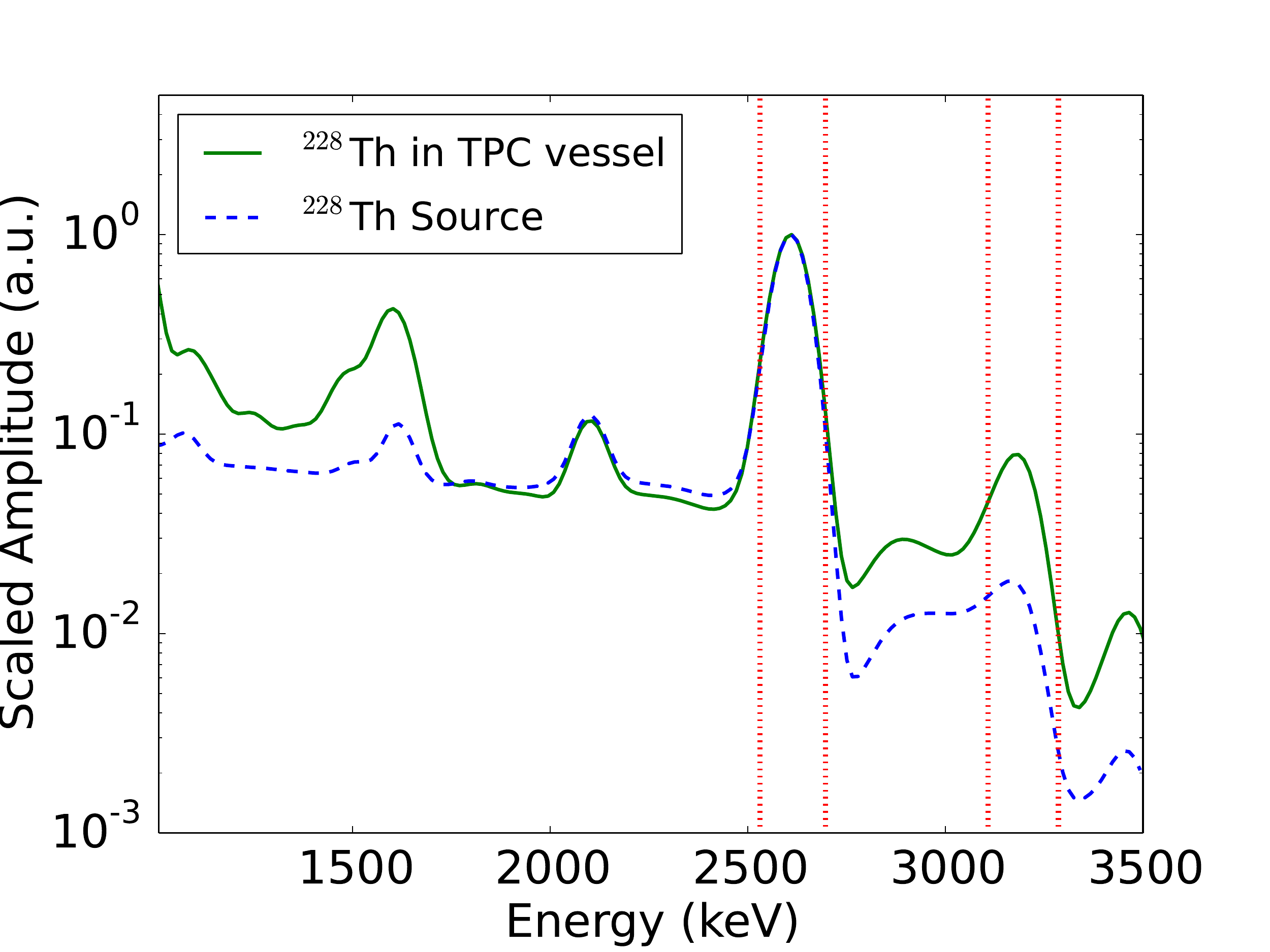}
\caption{(Color online) Comparison of MS PDFs for \isot{Th}{232} in the Cu TPC
vessel and a \isot{Th}{228} source located just outside the TPC at the cathode.
The PDFs are normalized to their respective 2615~keV peaks.  The difference in
spectral shapes arises from the energy-dependent attenuation length of
$\gamma$s and because some features are created by the coincident detection of
2 or more $\gamma$s.  The dotted (red) vertical lines indicate the energy
regions over which counts are integrated to form the ratio between the 2615~keV
$\gamma$ peak and the summation peak at 3197~keV(see text).
}
\label{fig:ThCompare}
\end{figure}

\begin{figure}
\includegraphics[width=0.98\columnwidth]{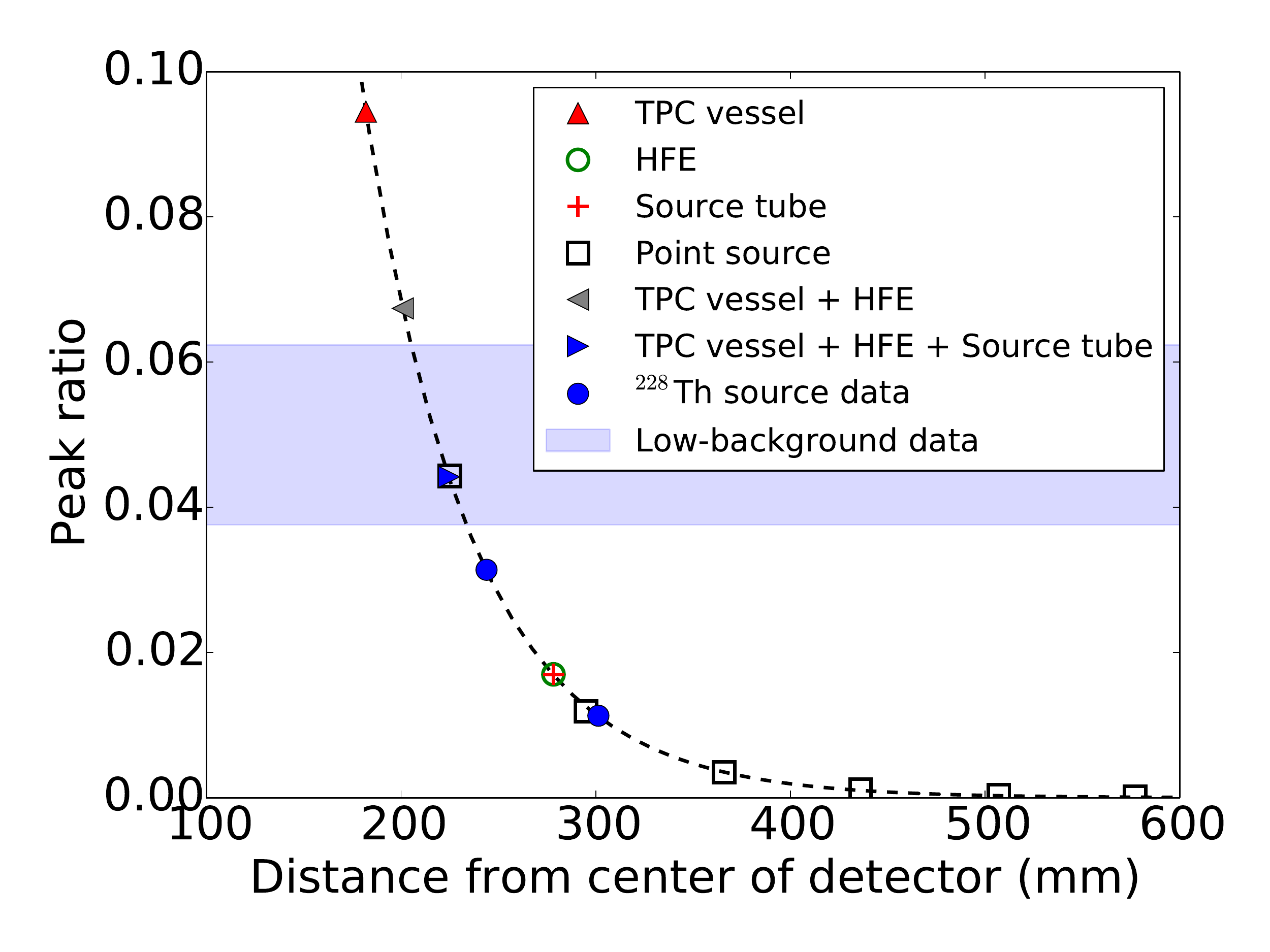}
\caption{(Color online) In the \thch{} spectrum, ratios of the summation peak
at 3197~keV and the 2615~keV $\gamma$ peak for low-background and
\isot{Th}{228} source
data, and from simulation of \thch{} point sources and \thch{} distributed in
particular detector components (see legend).  Note, the
following two pairs of points overlap: (1) `HFE' and `Source tube', and (2)
`TPC vessel + HFE + Source tube' and a `Point source' at x\cussim{220~mm}.  The
dashed line is an empirical fit to the simulated
point sources (hollow black squares) with the form $e^{a + b x}$, where $a$ and $b$ are fitted constants.
Results from source data (blue filled circles) agree well with it.  The fit is
used to interpolate the \emph{effective} distance from the center of the detector
for all non-point-like simulated geometries. 
Statistical error
bars on the points are all smaller than the size of the
symbols.  The shaded region is from the measured value in data, where the range
is due to statistical error.  This region agrees best with the combination of
PDFs from \thch{} in the TPC vessel, in the Cu source tube, and distributed in
the HFE, indicating the presence of a \thch{} background source external to the
TPC vessel.  }
\label{fig:ThVsPosition}
\end{figure}

Analyzing the relative strengths of peaks in the energy spectrum can yield
information on the position of a particular background source.  This is because 
the $\gamma$ attenuation length varies versus energy and 
spectral features (e.g.\ summation peaks) depend upon the coincident detection
of 2 or more separate $\gamma$s (see e.g.~\cref{fig:ThCompare}).  We have used the latter to investigate the location
of \thch{} in detector materials by analyzing the ratio of the summation peak
at 3197~keV (from 2615 and 583~keV $\gamma$s) to the 2615~keV $\gamma$ peak in
MS data.  The ratio is constructed from the number of counts in each peak,
defined by integrating over the $\pm2\sigma$ range at their particular
energies.  The ratio is calculated for PDFs derived from MC for components at
different locations in and around the detector, as well as for hypothetical
point sources of \thch{} placed at varying perpendicular distances from the TPC
in the same plane as the center cathode.  The ratio for combined PDF components
is produced by taking their normalized average, e.g.:

\begin{equation}
R_{\text{combined}} = \frac{
	\sum_{i} \frac{
		N_{\text{3197}}^{i}
		}{
		N_{\text{total}}^{i}
		}
	}{
	\sum_{i} \frac{
		N_{\text{2615}}^{i}
		}{
		N_{\text{total}}^{i}
		}	
	}
\end{equation}

where $N_{\text{3197, 2615, total}}^{i}$ are the number of counts in the 3197~keV summation peak,
in the 2615~keV peak, and in the total spectrum, respectively, for the
$i$th PDF component.  This construction assumes equal contributions from each PDF.  The ratio is measured for the low-background data also
by integrating in the energy regions as specified \textendash{} non-\thch{}-related backgrounds at the relevant energies are negligible.  A comparison
between this number and results from MC is presented in
\cref{fig:ThVsPosition}.  It is clear from these results that a non-negligible
source of \thch{} background exists external to the TPC vessel.  Possible
locations of this source are, for example, distributed in the surrounding HFE and/or
in the Cu tube used for deployment of radioactive sources adjacent to the TPC.
This is consistent with results from material assay, which indicate up to
\cussim{2.3}~counts/yr in the \nonubb{} ROI may arise from \thch{} backgrounds
external to the TPC (see e.g.~\cref{tab:ExpBkgdCts}).  The uncertainty in the actual location(s) of the \thch{}
background(s) is taken into account by including the far-source \thch{} in the
background model used in the EXO-200 data analysis. 

\begin{figure}
\includegraphics[width=0.98\columnwidth]{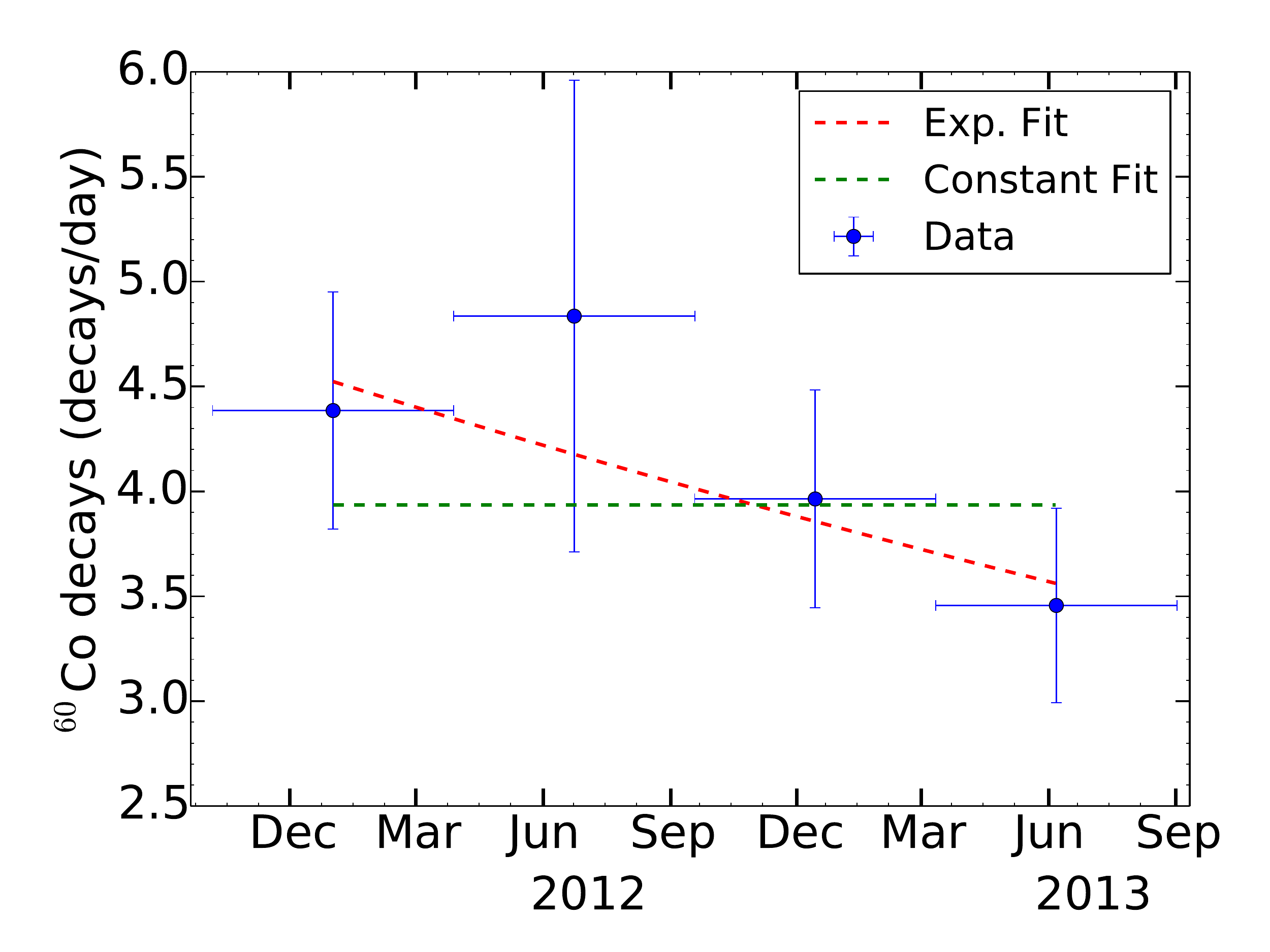}
\caption{(Color online) Behavior of \isot{Co}{60} fitted decay rate in the TPC vessel versus
time.  The rate shown is counts integrated above 980~keV.  (The full range used by the fit is 980~keV\textendash{}9800~keV, though the
\isot{Co}{60} PDF only extends to \cussim{2.5~MeV}). The vertical error bars on the
points are those returned by the fit to the low-background data
using the MIGRAD function of MINUIT~\cite{Bru97,Jam75}.  Horizontal error bars show the size of the time bins.  Two fits to the data
are shown, including a constant and an exponential fit.  The
exponential fit produces a half-life of $1500\pm1100 $~days,
consistent with the half-life of \isot{Co}{60}.  The $\chi^{2}/\text{NDF}$ is
0.5/2 (2.35/3) for the exponential (constant) fit.  } \label{fig:Co60VsTime}
\end{figure}

We have performed a similar study to investigate the possible presence of additional
\uch{} near the detector outside the TPC vessel.  The possibility of a
contribution from remote \uch{} has been taken into account in the background
model through the inclusion of \isot{Bi}{214} (resulting from \isot{Rn}{222} decays) in the air gap between the
cryostat and Pb shield.  The compatibility of this with measured Rn concentrations in the air
is discussed later.  In particular, we considered two additional far-source
components of \uch{} (distributed in the HFE and in the external Cu cryostat)
and add them, one at a time, to the background model to be fit to
the low-background data.  The (anti-)correlation in the fit between the
one additional remote background and the already included air gap \isot{Bi}{214} was
then determined.  The results of the two fits indicate a strong
anti-correlation between each pair of \uch{}-like PDFs.  In addition, there was
no statistical evidence in either of the two fits to demonstrate the presence
of \emph{both} \uch{}-like components in the data; in other words, the number of counts in
one of the PDFs in each pair was always consistent with zero.  Note
that this does not exclude the possibility of a \uch{} presence in the HFE or
in the Cu cryostat, but rather indicates the degeneracy of the PDFs and the
subsequent inability of the fit to distinguish between them.

\subsection{Time dependencies}

To look for time dependencies in the data, the low-background data were divided
into four time bins with roughly equal live-time.  The data in each time bin
were fit using the previously described procedure, and the number of counts
for each PDF extracted from the fit were tracked versus time.  Minor variations
over time are seen in the counts for some PDFs, but these can be
explained as either due to
statistical fluctuations or as (anti-)correlation between PDFs of similar shape
(e.g.\ \thch{} in the TPC vessel and far-source \thch{}).  We also investigated, in
particular, the behavior of the \isot{Co}{60} PDF in the TPC vessel, since the
total time range of data (\cussim{710~d}) is comparable to the \isot{Co}{60}
half-life ($1925.28\pm0.14$~d)~\cite{Browne20131849}.
Results for this particular PDF are shown in \cref{fig:Co60VsTime}.
Exponential fits to these data produce values consistent with the published
half-life, although with the present statistics a constant rate, such as
expected for continuous production, also provides a good fit. Estimates from
simulation suggest the generation of \isot{Co}{60} through
cosmic-ray-muon-produced neutrons to be negligible at WIPP depth. Further detector
exposure will improve the statistical power of this analysis.  

\subsection{\texorpdfstring{\isot{Xe}{137}}{Xe-137}}

\begin{figure}
\includegraphics[width=0.98\columnwidth]{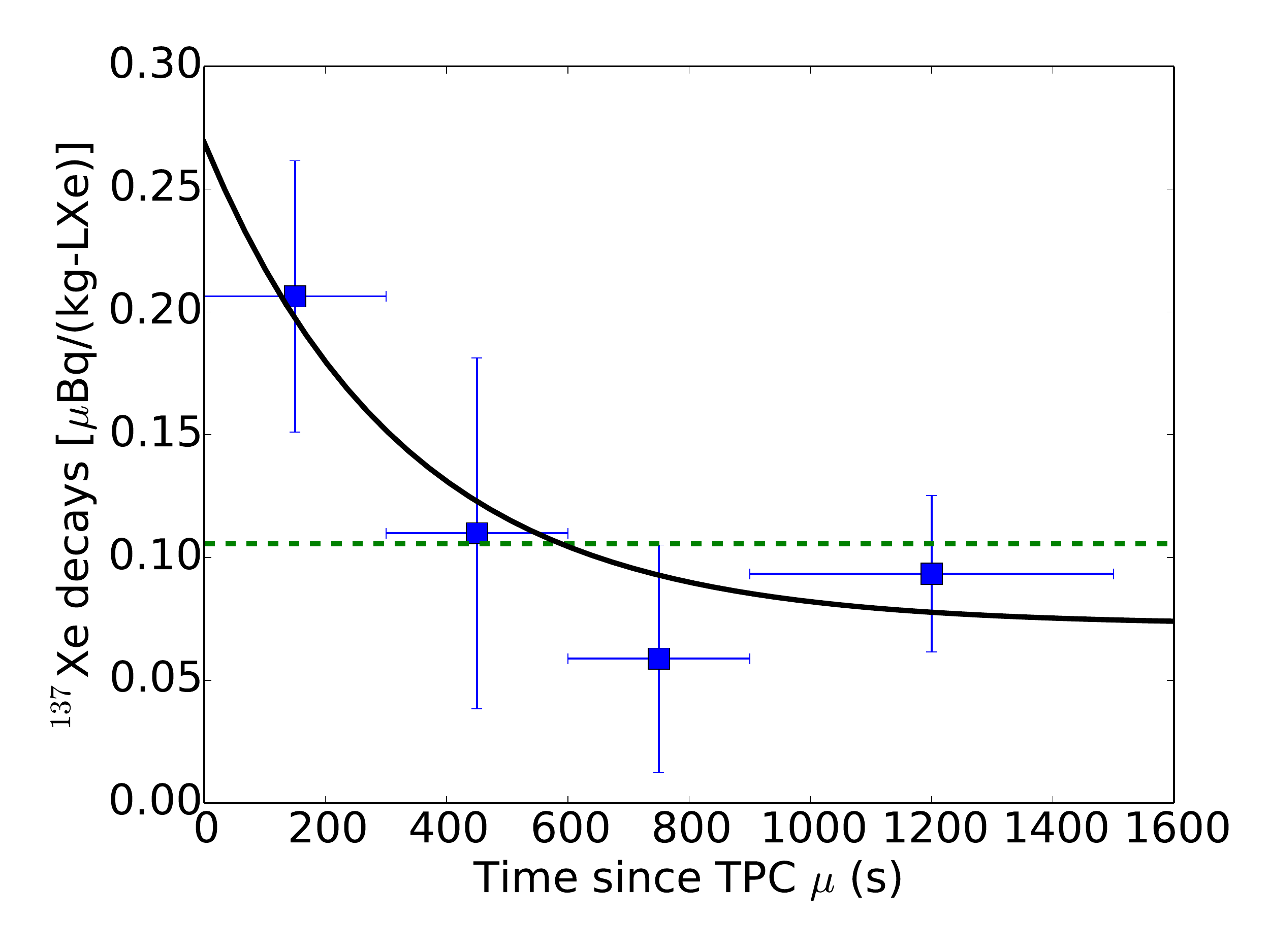}
\caption{(Color online) \isot{Xe}{137} decays 
versus time \emph{following} a muon passing through the TPC.  The horizontal
error bars indicate the width of the time bins used.  The data were fit to two
functions: (1) exponential decay + constant function, with the decay fixed to
the known \isot{Xe}{137} half-life, and (2) a constant.  The
$\chi^2$/NDF for 1 (2) is 0.84/2 (4.5/3), indicating a slight preference for 1. 
}
\label{fig:Xe137VsTime}
\end{figure}

Fits to the low-background data have indicated the presence of
\isot{Xe}{137} in the LXe~\cite{Nature2014}, a species that $\beta$-decays
(Q=$4173\pm7$~keV) with a half-life of $229.1\pm0.8$~s \cite{Browne20072173}
and which may be produced by neutron capture on \isot{Xe}{136}.  The ground-state decay, populated with a 67\% branching ratio,
produces a smooth spectrum of SS events up to its Q-value and is
predominantly responsible for SS events observed above the \isot{Tl}{208}
2615~keV peak.  We have searched for additional evidence of \isot{Xe}{137}
production following the transit of cosmic-ray muons through the TPC.  Such
muons (hereafter `\Tmus{}') coincide with higher neutron flux through the TPC and
therefore a higher expected neutron-capture rate on \isot{Xe}{136}.  They are efficiently
tagged owing to the large production of light and charge in the detector.
Normally the time following such events is removed from the low-background data
set because of the higher probability of producing short-lived radioactive
species in the LXe.  For this study, this selection requirement was relaxed and
low-background data were binned versus time following a \Tmu{}. The spectral
data in each time bin were fit using the standard procedure, and the behavior
of the \isot{Xe}{137} PDF was tracked versus time since \Tmu{}. The results of
this study are shown in \cref{fig:Xe137VsTime}.  The data were fit with an
exponential decay + constant function (decay constant fixed to the known
\isot{Xe}{137} half-life) as well as with simply a constant.  The $\chi^2$/NDF
indicates a slight preference for the decaying function, although the results of this decay-time study
are statistically still too weak to serve as direct evidence for \isot{Xe}{137}
production following a \Tmu{}.

\begin{figure}
\includegraphics[width=0.98\columnwidth]{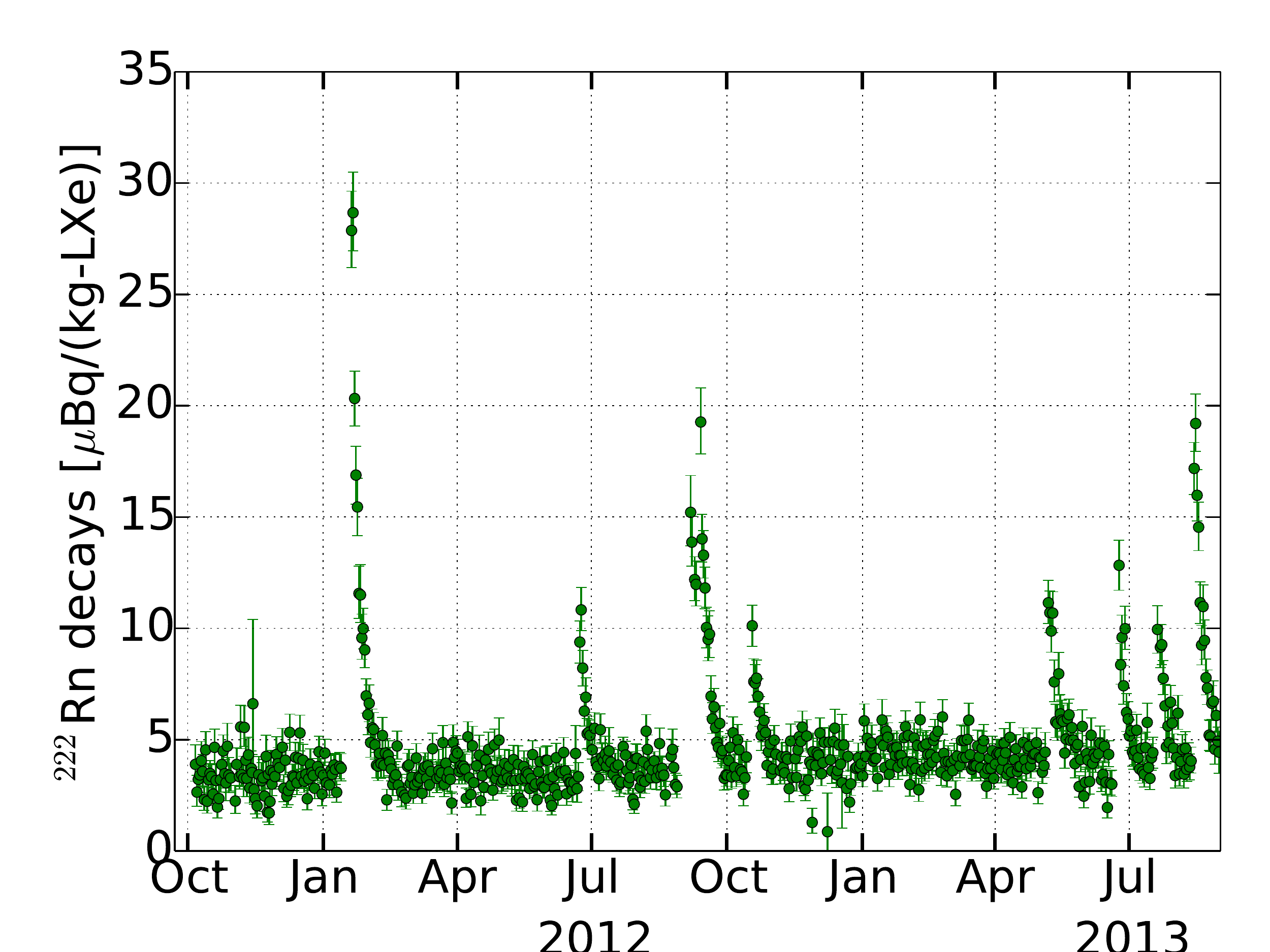}
\caption{(Color online) \isot{Rn}{222} decay rate in the LXe over \RunTwo{},
measured by tracking the counts of the 5.5~MeV decay~$\alpha$.  Spikes in the
count rate occur when the recirculation system is upset (typically by power
outages), resulting in the feed of some xenon from the bottle rack.   Such
xenon contains a higher amount of \isot{Rn}{222} (presumably from components in
the feed plumbing system) which then decays in the detector with the expected
half-life.  These periods of increased \isot{Rn}{222} activity are removed from
the low-background data set.}
\label{fig:RnRateTime}
\end{figure}

\subsection{Internal \texorpdfstring{\isot{Rn}{222}}{Rn-222}}

\isot{Rn}{222} internal to the TPC (e.g.~in the LXe and on the
cathode) can result in background counts in the \nonubb{} ROI from the decay of
\isot{Bi}{214}, which has a $\gamma$ line (2448~keV) near the
$\beta\beta$ Q-value.  We continuously monitor the actual amount of
\isot{Rn}{222} within the LXe by tracking counts of the 5.5~MeV decay $\alpha$ (see \cref{fig:RnRateTime}).
Energy depositions from $\alpha$s in the LXe are easily tagged using their
distinctive charge-light ratio.  In previous analyses, we find a rate of
$3.65\pm0.37$~\textmu{}Bq/(kg-LXe) corresponding to a steady-state population
of \cussim{200} \isot{Rn}{222} atoms~\cite{Albert2013}.  There are three
\isot{Rn}{222}-related components that may contribute background, all arising from decays of \isot{Bi}{214}: (a)
in the active LXe volume, (b) on the cathode, and
(c) in the inactive LXe volume.  The decay of \isot{Rn}{222} and
its daughters in the active LXe sometimes create positively-charged ions that drift
towards the cathode.  Only 17\% of the resulting \isot{Bi}{214} decays occur in
the active xenon (a), with the remaining 83\% occurring on the cathode
(b)~\cite{Albert2013}.  A majority of the (a) and (b) decays ($>50\%$) are tagged using
\isot{Bi}{214}\textendash{}\isot{Po}{214} decay coincidences
(T$_{1/2}^{^{214}{\mathrm{Po}}} = 163.6\pm0.3$~\mus{}~\cite{ShamsuzzohaBasunia2014561}).  In contrast, decays in
the inactive LXe (c) are not actively tagged.  We calculate the contribution of
untagged events to the \nonubb{} ROI using MC and present the results in
\cref{tab:IntRn}.  The contribution of internal-Rn-related backgrounds to the
\nonubb{} ROI is insignificant in comparison to those from e.g.~external \thch{} and
\uch{} sources. 

\begin{table} 
\renewcommand{\arraystretch}{1.3}
\caption{Background contributions (untagged) from \isot{Rn}{222} internal to the TPC
vessel (inside the LXe and on the cathode) estimated using low-background data
and simulation.  The fiducial volume is that used in~\cite{Nature2014},
corresponding to an active (inactive) LXe mass of 94.6~kg (61.7~kg). Errors on
each rate are $\pm10\%$.} 
\label{tab:IntRn} 
\begin{ruledtabular}
\begin{tabular}{lc}
    \toprule
   	\isot{Rn}{222} component & \multicolumn{1}{m{0.47\columnwidth}}{\centering Expected counts in \nonubb{} ROI (Cts / yr)} \\
    \colrule
	\isot{Rn}{222} in the active LXe & 0.011 \\
	\isot{Bi}{214} on the cathode & 0.14    \\
	\isot{Rn}{222} in the inactive LXe & 0.085\\
    \colrule
    Total: & 0.24 \\
\end{tabular}
\end{ruledtabular}
\end{table}

The measurement of a constant rate of \isot{Rn}{222} decay (except for the spikes discussed in the \cref{fig:RnRateTime} caption) indicates that its
population is supported, likely from the emanation from materials in contact
with the LXe.  This may occur either outside or inside the TPC, as the LXe is
continuously circulated and purified during normal detector operation.
Measurements of Rn emanation from the external xenon piping show that
contributions from this system may explain the steady-state population of Rn in
the LXe.  For components internal to the TPC, outgassing estimates based on the
measured U content (or lack thereof) indicate that only the internal polyimide readout
cables (with a surface area of 37600~cm$^2$ in contact with
LXe) could provide a significant contribution to the Rn population.  Rn
emanation measurements of these cables only provide upper limits on this
contribution, demonstrating they are capable of supporting up to 270
\isot{Rn}{222} atoms in the LXe. 

\subsection{External \texorpdfstring{\isot{Rn}{222}}{Rn-222}}

An earlier search for \nonubb{} using a smaller data set
(e.g.~\cite{Auger:2012ar}) showed an indication for an background
source of \uch{} outside of the TPC vessel, possibly from \isot{Rn}{222} in the
air gap between the Cu cryostat and the inside of the Pb shield.  The \isot{Rn}{222}
content in the cleanroom air was monitored during most of the \RunTwo{}
data-taking period using a commercial Rad7 device~\cite{Rad7}.  Pressure
measurements in the air gap indicate a complete exchange with the cleanroom air
every 6 minutes, implying that the \isot{Rn}{222} content of both air volumes
should be in equilibrium.  Measurements with the Rad7 device produce an average
\isot{Rn}{222} decay rate per unit volume of $6.6\pm0.3$~Bq/m$^{3}$ over the
\RunTwo{} time period.  In addition, diurnal and annual modulations in the
\isot{Rn}{222} decay rate were observed, both exhibiting peak-to-peak
variations of \cussim{4~Bq/m$^3$}.

We analyze low-background data from a subset of \RunTwo{} (ending in June 2013)
and perform a fit to this data, using the standard procedure described in
\cref{sec:Intro}.  Using the efficiency calculated from MC, the volume in the
air gap, and assuming that the far \uch{}
is arising entirely from \isot{Rn}{222} in the air gap, we obtain the estimated
activity of \isot{Rn}{222} in this space: \pmasy{89}{18}{27}~Bq/m$^{3}$.  This
value is inconsistent with the Rad7 measurements, strongly suggesting that
far-source \uch{}-like contributions are not predominantly from \isot{Rn}{222}
in the air gap, but rather from other sources including the HFE or the Cu
cryostat.  Such contributions are still allowed by limits from material assays,
which estimate a contribution from external \uch{} backgrounds in the
\nonubb{} ROI of up to \cussim{1.5}~counts/yr.  Additional analyses have searched for
annual and diurnal modulations of this background component, by (1) binning
versus time, and (2) binning the data according to time of day and fitting the
spectral data of each bin.  Neither search finds a statistically
significant modulation. 

\section{Conclusions}

The results presented here give confidence in the background model used in
previous publications with data from the EXO-200 detector.  Whereas 
some information may be obtained from the low-background data concerning the
location of background sources, it is difficult to determine exact positions
because of the overall low background count rate.  The EXO-200 detector and
related facilities at WIPP are currently undergoing preparations for
re-commissioning in 2015, following an extended data-taking hiatus owing to the
closure of the WIPP underground beginning early 2014.  The planned deployment
of a charcoal-based Rn-suppression system should help reduce the \isot{Rn}{222} content in the air
gap inside the Pb shield by a factor of 10 -- 100, though current results suggest that this may not have
a significant impact on the observed background.  At the minimum, this will
reduce \isot{Rn}{222} to such a level so as to allow one degenerate fit
component to be removed from the background model.  Development of analysis
techniques to reduce \isot{Xe}{137} backgrounds in the data are being explored,
with the goal to apply them to future and existing data sets from the
EXO-200 detector. 

The agreement observed between background estimates from material assay and
results from fits of the EXO-200 data provides a basis for employing similar
calculations for the nEXO detector, a proposed 5000~kg next-generation \nonubb{} experiment using LXe and based on the EXO-200 design.  The analyses
and comparisons presented in this paper show that careful component radio assay
is able to produce both robust estimations of experiment sensitivity as well as
provide guidance and constraint for its design.  In addition, cosmogenic
backgrounds for the nEXO detector, including \isot{Xe}{137}, will be
significantly reduced by deploying to a deeper location where the cosmic-ray muon
flux is at least two orders of magnitude lower than at WIPP.  Estimates using
simulation for \isot{Xe}{137} production at deep sites ($>6$~km water
equivalent) indicate a reduction of at least three orders of magnitude with
comparison to the WIPP site.

\begin{acknowledgments}
	EXO-200 is supported by DOE and NSF in the United States, NSERC in Canada,
SNF in Switzerland, IBS in Korea, RFBR (14-02-00675) in Russia, CAS-IHEP Fund in China, and DFG Cluster
of Excellence ``Universe'' in Germany. EXO-200 data analysis and simulation
uses resources of the National Energy Research Scientific Computing Center
(NERSC), which is supported by the Office of Science of the U.S. Department of
Energy under Contract No.~DE-AC02-05CH11231. The collaboration gratefully
acknowledges the WIPP for their hospitality.

\end{acknowledgments}

\bibliography{exo_backgrounds}

\end{document}